\documentclass[aps,prl,twocolumn,groupedaddress,showpacs]{revtex4}
\usepackage{graphicx}
\usepackage{helvet}

\begin{document}

\title{Spin Liquid State in an Organic Mott Insulator with Triangular Lattice}

\author{Y. Shimizu,$^{1,2}$ K. Miyagawa,$^{2}$ 
	K. Kanoda,$^{2,3}$ M. Maesato,$^{1}$ and G. Saito$^{1}$
}
	
\affiliation{
$^{1}$ Division of Chemistry, Graduate School of Science, Kyoto University, Oiwaketyo, Kitashirakawa, Sakyo-ku, Kyoto, 606-8502, Japan \\
$^{2}$ Department of Applied Physics, University of Tokyo, Hongo, Bunkyo-ku, Tokyo, 113-8656, Japan. \\
$^{3}$ CREST, Japan Science and Technology Corporation (JST). \\}

\date{\today}

\begin{abstract}
	$^{1}$H NMR and static susceptibility measurements have been performed in an organic Mott insulator with nearly isotropic triangular lattice, $\kappa$-(BEDT-TTF)$_{2}$Cu$_{2}$(CN)$_{3}$, which is a model system of frustrated quantum spins. The static susceptibility is described by the spin $S$ = 1/2 antiferromagnetic triangular-lattice Heisenberg model with the exchange constant $J$ $\sim$ 250 K. Regardless of the large magnetic interactions, the $^{1}$H NMR spectra show no indication of long-range magnetic ordering down to 32 mK, which is four-orders of magnitude smaller than $J$. These results suggest that a quantum spin liquid state is realized in the close proximity of the superconducting state appearing under pressure.
\end{abstract}



\maketitle

\begin{figure}
\includegraphics[height=4cm, width=4cm]{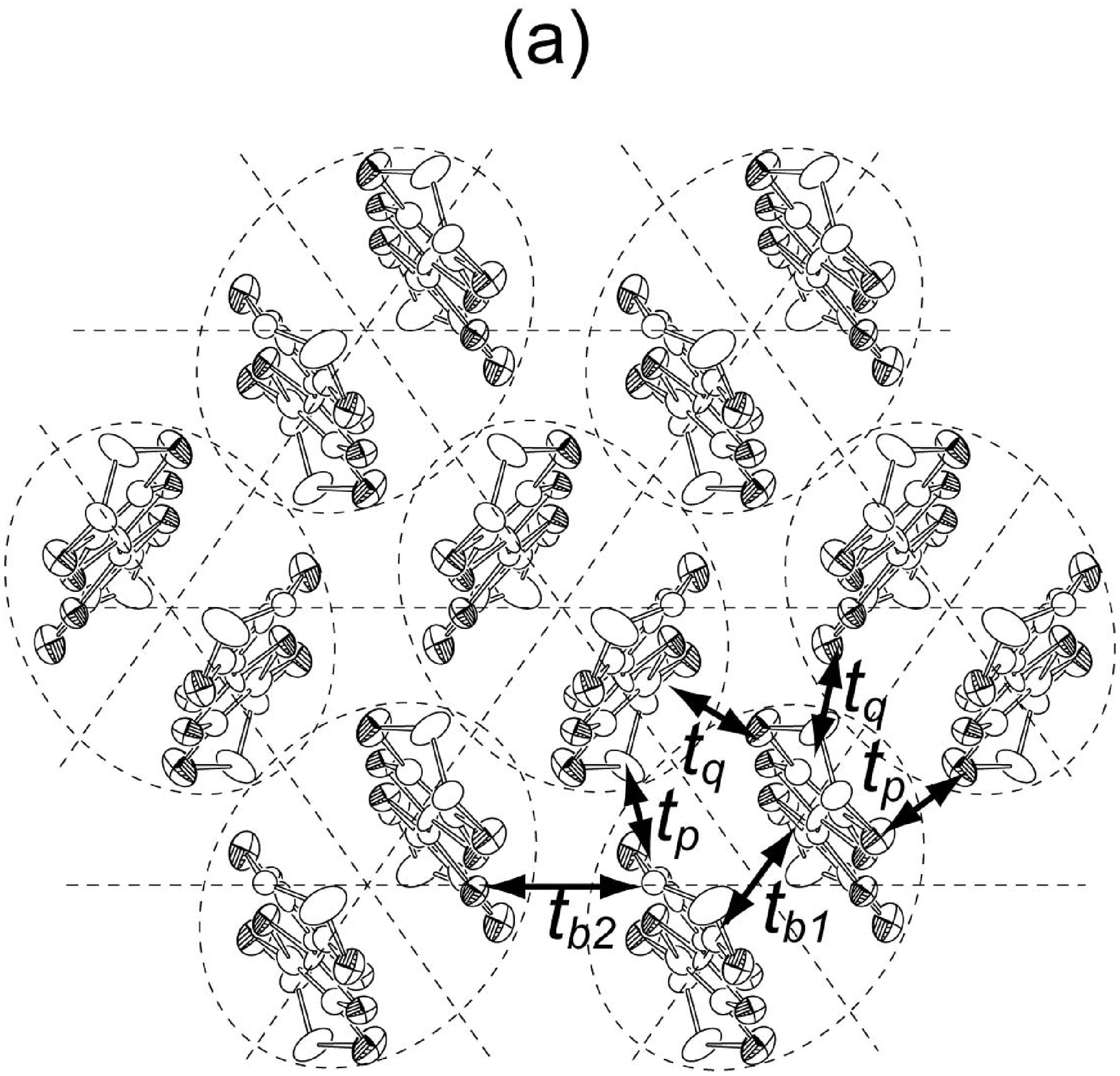}
\includegraphics[height=4cm, width=4cm]{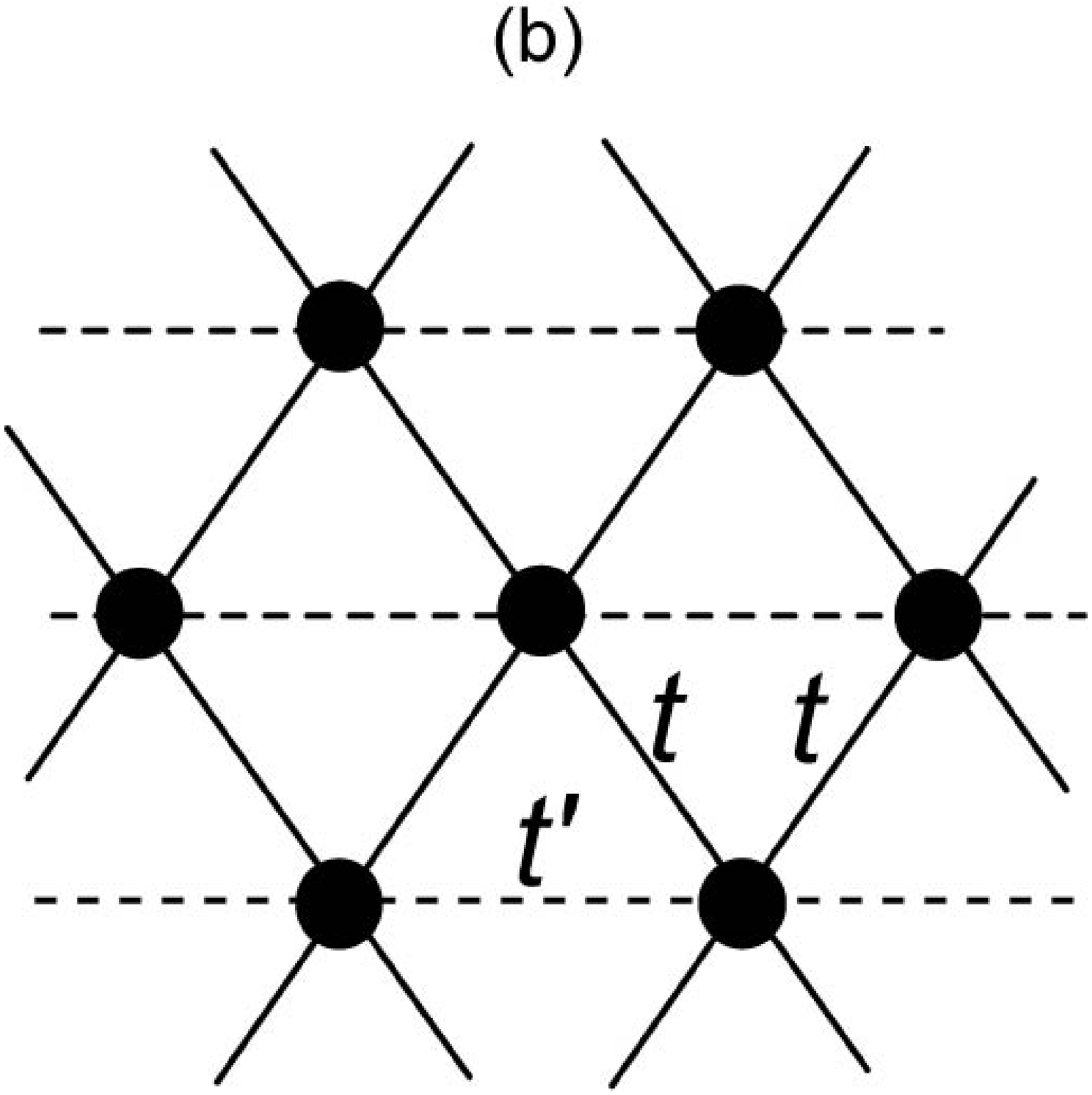}
\caption{\label{Fig1}(a) Crystal structure of an ET layer of $\kappa$-(ET)$_{2}$Cu$_{2}$(CN)$_{3}$ viewed along the long axes of ET molecules [6]. The the transfer integrals between ET molecules, $t_{\rm b1}$, $t_{\rm b2}$, $t_{\rm p}$ and $t_{\rm q}$, are calculated as 22 meV, 11 meV, 8 meV and 3 meV, respectively [7]. For the large $t_{\rm b1}$ compared with other transfer integrals, the face-to-face pair of ET molecules connected with $t_{\rm b1}$ can be regarded as dimer unit consisting of the triangular lattice. (b) Schematic representation of electronic structure of $\kappa$-(ET)$_{2}$X, where the dots represent the ET dimer units. They form the anisotropic triangular lattice with $t = (|t_{\rm p}| + |t_{\rm q}|)/2$ and $t^{\prime} = t_{\rm b2}/2$.}\end{figure}

 The magnetism of the Mott insulator, which is the mother phase giving the unconventional superconductivity in the high-$T_{\rm C}$ cuprates and $\kappa$-(BEDT-TTF)$_{2}$X organics, has been attracting much attention, because it holds the key to understand the mechanism of the superconductivity, where BEDT-TTF (ET) denotes bis(ethylenedithio)-tetrathiafulvalene and X denotes inorganic mono-valent anion [1,2]. The ground states of the Mott insulators studied so far in these materials are antiferromagnets. The stage of the interacting spins are quasi two-dimensional square lattice or anisotropic triangular lattice with the nearest neighbor transfer $t$ and the second-nearest neighbor transfer $t^{\prime}$. If the lattice is close to isotropic triangle ($t^{\prime}/t$ $\sim$ 1), however, the geometrical frustration gets to work significantly against the long-range magnetic ordering (LRMO), and a spin liquid state without symmetry breaking, which attracts great interest as an exotic state, can emerge [3].

 In the case of $\kappa$-(ET)$_{2}$X, dimerization of a face-to-face ET pair is strong enough to treat the dimer as an unit [Fig. 1(a)], and the system can be effectively described by Hubbard model on an anisotropic triangular lattice [Fig. 1(b)] with a half-filled conduction band [4,5]. The effective transfer integrals between the dimers are given as $t = (|t_{\rm p}| + |t_{\rm q}|)/2$ and $t^{\prime} = t_{\rm b2}/2$, respectively, where $t_{\rm p}$, $t_{\rm q}$ and $t_{\rm b2}$ are transfer integrals shown in Fig.1(a) and evaluated with the extended H\"{u}ckel method and the tight-binding approximation. Among the $\kappa$-(ET)$_{2}$X family, a Mott insulator $\kappa$-(ET)$_{2}$Cu$_{2}$(CN)$_{3}$ [6,7] is unique in that the ratio of transfer integrals is almost unity ($t^{\prime}/t$ = 1.06) [5], suggesting that the $S$ = 1/2 nearly isotropic triangular lattice is realized and it can be a promising candidate of the spin liquid insulator. Actually, the EPR measurement has shown no signature of the antiferromagnetic (AF) transition down to 1.7 K [7], although the nature of the spin state is still unknown. It is in sharp contrast to another Mott insulator $\kappa$-(ET)$_{2}$Cu[N(CN)$_{2}$]Cl with $t^{\prime}/t$ $\sim$ 0.75, which exhibits the AF transition at $T_{\rm N}$ = 27 K at ambient pressure [8,9] and the superconducting transition at $T_{\rm C}$ = 12.8 K under pressure [10]. It is also noted that moderate hydrostatic pressure induces superconductivity in $\kappa$-(ET)$_{2}$Cu$_{2}$(CN)$_{3}$ with $T_{\rm C}$ of 3.9 K [7].

 In this letter, we report the magnetic properties of $\kappa$-(ET)$_{2}$Cu$_{2}$(CN)$_{3}$ revealed by the $^{1}$H NMR and the static susceptibility measurements. We have observed no LRMO down to 32 mK well below the exchange constant $J$ = 250 K estimated from the magnetic susceptibility at ambient pressure and $T_{\rm C}$ under soft pressure. These results strongly suggest that a quantum spin liquid state is likely realized in the neighborhood of the superconducting phase.

 The single crystals of $\kappa$-(ET)$_{2}$Cu$_{2}$(CN)$_{3}$ were prepared by the standard electrochemical method [6,7]. The magnetic susceptibility was measured for a polycrystalline sample in a temperature range from 1.9 K to 298 K at 0.32 T. The $^{1}$H NMR experiments were performed for a polycrystalline sample in a temperature range of 1.4 - 200 K at a field of 3.9 T and for a single crystal weighing 76 $\mu$g in a range of 32 mK - 36 K at 2.2 T applied normal to the conducting plane. The latter measurements were performed using the dilution refrigerator of the top-loading type with the crystal soaked to the $^{3}$He-$^{4}$He mixture. The absence of Cu$^{2+}$ impurity ($< 0.01\%$) was confirmed by EPR before the $^{1}$H NMR measurement. The NMR spectra were obtained by the fast Fourier transformation of the quadrature-detected echo signals. The relaxation curves of nuclear magnetization were obtained from the recovery of the echo intensity following saturation comb pulses and the solid-echo pulse sequence, $(\pi/2)_{\rm x}-(\pi/2)_{\rm y}$.

\begin{figure}
\includegraphics[height=8.5cm, width=7cm]{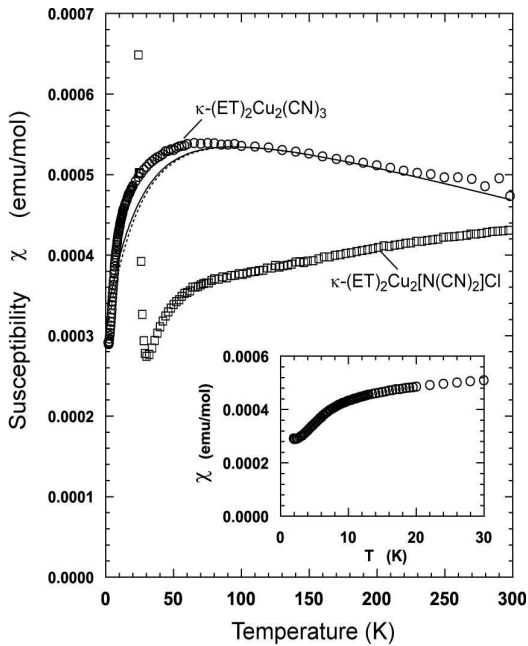}
\caption{\label{Fig2}Temperature dependence of the magnetic susceptibility of the randomly orientated polycrystalline samples of $\kappa$-(ET)$_{2}$Cu$_{2}$(CN)$_{3}$ and $\kappa$-(ET)$_{2}$Cu[N(CN)$_{2}$]Cl [9]. The core diamagnetic susceptibility is already subtracted. The solid and dotted line represent the result of the series expansion of triangular-lattice Heisenberg model using [6,6] and [7,7] Pad$\acute{e}$ approximants, respectively, with $J$ = 250 K. The low-temperature data of $\kappa$-(ET)$_{2}$Cu$_{2}$(CN)$_{3}$ below 30 K are expanded in the inset.}
\end{figure}

 Temperature dependence of the static susceptibility, $\chi$, of $\kappa$-(ET)$_{2}$Cu$_{2}$(CN)$_{3}$ is shown in Fig. 2, where the core diamagnetic contribution of - 4.37 $\times$ 10$^{-4}$ emu/mol is already subtracted. With decreasing temperature, $\chi$ increases slightly and shows a very broad maximum around 70 K (5.4 $\times$ 10$^{-4}$ emu/mol). Below 50 K, $\chi$ starts to decrease rapidly, but remains to be paramagnetic even at 1.9 K (2.9 $\times$ 10$^{-4}$ emu/mol). The behavior is quite different from that of $\kappa$-(ET)$_{2}$Cu[N(CN)$_{2}$]Cl which shows a monotonous decrease with temperature and the weak-ferromagnetism below 27 K due to canting of the AF ordered spins [9]. The temperature dependence of $\chi$ for $\kappa$-(ET)$_{2}$Cu$_{2}$(CN)$_{3}$ is fitted to the high-temperature series expansion of spin $S$ = 1/2 triangular lattice Heisenberg model [11] as shown in Fig. 2, where the [6/6] and [7/7] Pad$\acute{e}$ approximants are adopted with $J$ = 250 K. This model was successful in explaining $\chi$ of another organic triangular lattice system [12]. The peak temperature is much lower than the $J$ value, suggesting that the strong spin frustration suppresses the development of the short-range spin correlations. The difference between the experimental result and the Heisenberg model may be partially attributed to the weak spin localization in the present system situated in the vicinity of the Mott transition.

\begin{figure}
\includegraphics[height=7cm, width=8cm]{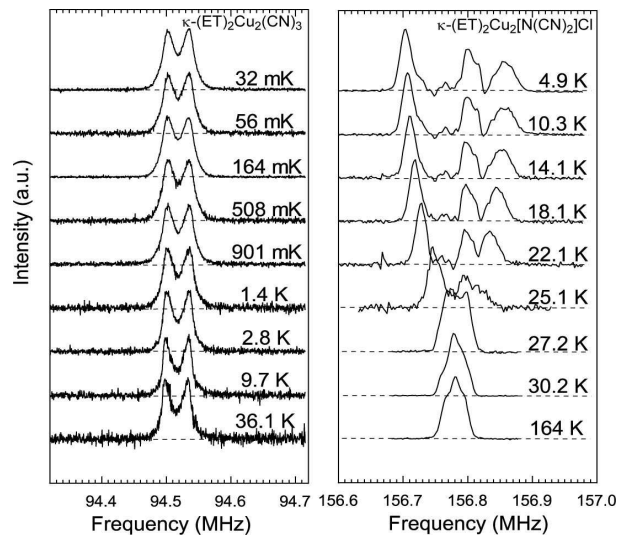}
\caption{\label{Fig3}(a) $^{1}$H NMR absorption spectra for single crystals of $\kappa$-(ET)$_{2}$Cu$_{2}$(CN)$_{3}$ and $\kappa$-(ET)$_{2}$Cu[N(CN)$_{2}$]Cl [9] under the magnetic field perpendicular to the conducting planes.}.
\end{figure}

 Figure 3 shows the temperature dependence of the $^{1}$H NMR spectra of a single crystal of $\kappa$-(ET)$_{2}$Cu$_{2}$(CN)$_{3}$ along with the previous result of $\kappa$-(ET)$_{2}$Cu[N(CN)$_{2}$]Cl for comparison [9]. The width and the shape of the spectra of the both salts above 30 K represent typical nuclear dipole interactions between the protons in the ethylene groups of ET molecules. Since the shape of the spectra is sensitive to the direction of the external static magnetic field, the difference of the spectra between the two salts at high temperatures is explained by the difference in the orientation of ET molecules against the applied field and does not matter. A remarkable difference in the shape of the spectra between the two salts was observed at low temperatures. The spectra of $\kappa$-(ET)$_{2}$Cu[N(CN)$_{2}$]Cl clearly split below 27 K and the splitting width reaches $\pm$ 80 kHz, reflecting the commensurate AF ordering [9] with a magnetic moment of 0.45$\mu_{\rm B}$ per an ET dimer [13]. On the other hand, the spectra of $\kappa$-(ET)$_{2}$Cu$_{2}$(CN)$_{3}$ show neither distinct broadening nor split down to 32 mK. The result indicates that no LRMO exists in $\kappa$-(ET)$_{2}$Cu$_{2}$(CN)$_{3}$ at least down to 32 mK, which is four-orders of magnitude below the $J$ value of 250 K. The fact strongly suggests the realization of the quantum-disordered spin liquid state in $\kappa$-(ET)$_{2}$Cu$_{2}$(CN)$_{3}$ due the strong spin frustration of the nearly isotropic triangular lattice. Taking a closer look at the data, the full width of the spectra at the half-maximum intensity shows a slight broadening of $\pm$ 2 kHz with decreasing temperature from 4 K to 1 K. It may originate from the random dipole field of a small amount of magnetic impurity or the intrinsic $T_{2}$-broadening as was observed in the triangular lattice compound, LiNiO$_{2}$ [14], where $T_{2}$ is the spin-spin relaxation time. The magnetic moment, if any below 4 K, is estimated as less than 0.01$\mu_{\rm B}$ per an ET dimer with reference to the moment/shift ratio observed in $\kappa$-(ET)$_{2}$Cu[N(CN)$_{2}$]Cl.

 The nuclear spin-lattice relaxation rate, $T_{1}^{-1}$, of $\kappa$-(ET)$_{2}$Cu$_{2}$(CN)$_{3}$ is shown in Fig.4 as a function of temperature together with that of $\kappa$-(ET)$_{2}$Cu[N(CN)$_{2}$]Cl [9]. An enhancement of $T_{1}^{-1}$ above 150 K is a motional contribution due to the thermally activated vibration of the ethylene groups. The motional contribution almost dies away around 150 K, below which the relaxation is electronic in origin. From 150 K to 50 K, $T_{1}^{-1}$ behaves nearly temperature-independent. The values of $T_{1}^{-1}$ in this region are more than twice as large as those of $\kappa$-(ET)$_{2}$Cu[N(CN)$_{2}$]Cl.

\begin{figure}
\includegraphics[height=10cm, width=8cm]{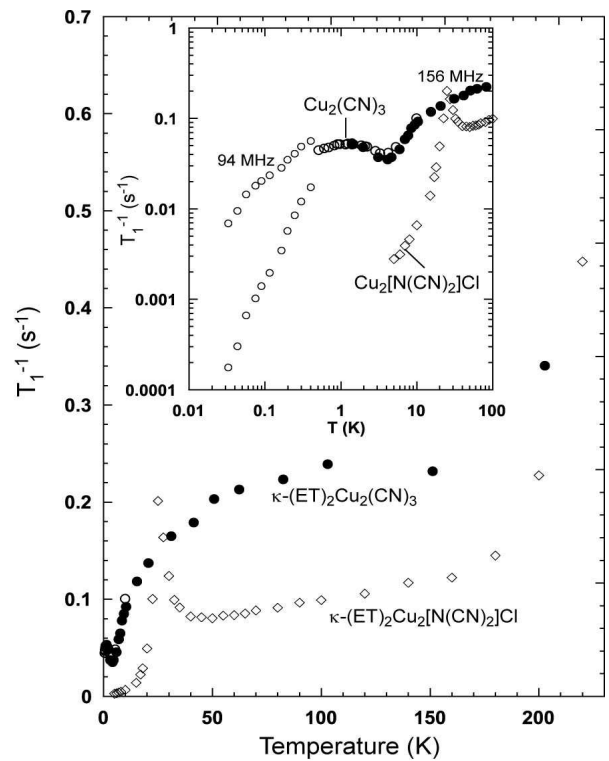}
\caption{\label{Fig4}$^{1}$H nuclear spin-lattice relaxation rate $T_{1}^{-1}$ above 1 K for a single crystal (open circles) and a polycrystalline sample (closed circles) of $\kappa$-(ET)$_{2}$Cu$_{2}$(CN)$_{3}$ and a single crystal of $\kappa$-(ET)$_{2}$Cu[N(CN)$_{2}$]Cl (open diamonds) [9]. The inset shows the data down to 32 mK in logarithm scales.}
\end{figure}

 Below 50 K, $T_{1}^{-1}$ of $\kappa$-(ET)$_{2}$Cu$_{2}$(CN)$_{3}$ decreases with temperature down to 4 K in a similar manner to $\chi$. It is seen that there is no difference between the polycrystalline and single crystal data in the overlapping temperature range of 1.4 K-36 K. This temperature dependence markedly contrasts with that of $\kappa$-(ET)$_{2}$Cu[N(CN)$_{2}$]Cl having a sharp peak at 27 K, which is characteristic of the magnetic transition. Since $\chi$ and $T_{1}^{-1}$ measure the $\bf{k}$ = $\bf{0}$ uniform component and the summation of the spin fluctuations in the $\bf{k}$-space, respectively, the results of $\chi$ and $T_{1}^{-1}$ suggest that the spin excitations are suppressed below 50 K over the $\bf{k}$-space. Below 4 K, however, $T_{1}^{-1}$ of $\kappa$-(ET)$_{2}$Cu$_{2}$(CN)$_{3}$ turns to increase and shows a broad peak around 1 K as shown in the inset in Fig. 4. It is noted that $\chi$ has no appreciable anomaly around 4 K, where $T_{1}^{-1}$ shows an upturn. The broad peak is considered to reflect the characteristic structure in the spin excitation spectrum of quantum liquid with slow spin dynamics.

 Below about 0.4 K the relaxation curve starts to bend gradually and fits to a sum of two exponential functions with {\it comparable} fractions. The temperature dependences of the two components of $T_{1}^{-1}$ are proportional to $\sim$ $T^{2}$ and $\sim$ $T$. It means that the two kinds of proton sites with different $T_1$ are separated in a macroscopic scale but not in a molecular scale, because the $T_1$ distribution in the molecular scale would be averaged by the $T_2$ process during such a slow spin-lattice relaxation as in the present case ($T_1$ $\sim$ $10^2$ sec and $10^4$ sec at the lowest temperature). It is unlikely that LRMO or spin glass transition occurs below 0.4 K, since $T_{1}^{-1}$ has no critical behavior above 0.4 K and the NMR spectra show no broadening below 1 K. Actually, the recent $\mu$SR experiment also shows no internal magnetic field down to 20 mK [15]. Taking account that the highly degenerate quantum state on the triangular lattice is likely sensitive to randomness [16], crystal imperfections or dilute impurity moments may induce some secondary phase with different spin dynamics from the primary phase. It is noted that the inhomogeneous NMR relaxation like the present observation is also encountered in an inorganic frustrated spin system with kagom$\acute{e}$ lattice [17]. In general, the spin liquid state is discussed to have a finite spin excitation gap [3]. In such a case, $T_{1}^{-1}$ should fall exponentially with lowering temperature. In the present case, however, $T_{1}^{-1}$'s follow power laws at low temperatures, indicating that the ground state has low-lying spin excitations of gapless nature.

 The ground state of the $S$ = 1/2 triangular lattice Heisenberg model has been intensively studied [18] since Anderson proposed the spin liquid state [3]. Now, there is an accepted consensus that the ground state of the isotropic triangular lattice is the $120^{\circ}$ spiral ordered state. The quantum-disordered state is suggested to appear when the triangular lattice becomes anisotropic [19], which is, in fact, realized in Cs$_2$CuCl$_4$ ($J^{\prime}/J$ $\sim$ 3) under high magnetic fields [20]. On the other hand, the path integral renormalization group analysis of the $\it{Hubbard}$ model has predicted that non-magnetic insulating state appears in a restricted region of the $U/t$ {\it vs}. $t^{\prime}/t$ plane with $\it{moderate}$ values of $U/t$ exceeding the critical value of the Mott transition and the region is widen as $t^{\prime}/t$ approaches unity [21]. Here $U$ represents the effective on-site Coulomb repulsive energy. The band structure calculation suggests that $\kappa$-(ET)$_{2}$Cu$_{2}$(CN)$_{3}$ ($U/t$ = 8.2, $t^{\prime}/t$ = 1.06) is just located in the predicted non-magnetic region. It is interesting that the frustrated Hubbard ladder model predicts the similar phase diagram in the $U/t$ {\it vs}. $t^{\prime}/t$ plane [22]. These theoretical works imply why the promising candidates of the spin liquid phase are realized on the nearly isotropic triangular lattice in organics, where $U/t$ is usually not so large. Actually, the fact that $\kappa$-(ET)$_{2}$Cu$_{2}$(CN)$_{3}$ undergoes the Mott transition under soft hydrostatic pressure [7] indicates the marginal $U/t$ at ambient pressure.

 Finally, it should be reminded that the superconducting phase appears in $\kappa$-(ET)$_{2}$Cu$_{2}$(CN)$_{3}$ with the maximum $T_{\rm C}$ of 3.9 K under moderate hydrostatic pressure. To our knowledge, this is the first example of Mott insulator which shows no LRMO well below the superconducting transition appearing under pressure or carrier doping. The superconducting and metallic phases emerging from the spin liquid insulator can have novel aspects unseen in the high-$T_{\rm C}$ oxides and the other organic analogues [1, 23] where the $\it{d}$-wave superconductors and the pseudo-gapped metals emerge from the AF insulators.

 In conclusion, through the $^{1}$H NMR study on the Mott insulator $\kappa$-(ET)$_{2}$Cu$_{2}$(CN)$_{3}$ with nearly isotropic triangular lattice, we found the evidences for the absence of LRMO down to 32 mK, despite of the large AF interaction of hundreds Kelvin. The result is quite different from the other Mott insulators such as $\kappa$-(ET)$_{2}$Cu[N(CN)$_{2}$]Cl and indicates that the spin liquid state is realized in the proximity of the superconducting phase under pressure.

 We thank M. Imada, H. Fukuyama, N. Nagaosa, Y. Kitaoka and K. Yoshimura for enlightening discussions. This work was supported by NEDO and Grant-in-Aids for COE Researches on "Elements Science" and "Phase Control in Spin-Charge-Photon Coupled Systems" from the Ministry of Education, Science, Sports and Culture of Japan, and for Scientific Research (No. 14204033) by JSPS.


\begin{thebibliography}{99}
\bibitem{Ref1} K. Kanoda, Physica C {\bf 282}-{\bf 287}, 299 (1997); 
K. Kanoda, Hyperfine Interact. {\bf 104}, 235 (1997)

\bibitem{Ref2} R. H. McKenzie, Science {\bf 278}, 820 (1997).

\bibitem{Ref3} P. W. Anderson, Mater. Res. Bull. {\bf 8}, 153 (1973).

\bibitem{Ref4} H. Kino and H. Fukuyama, J. Phys. Soc. Jpn. {\bf 64}, 2726 (1995).

\bibitem{Ref5} R. H. McKenzie, Comments Cond. Mat. Phys. {\bf 18}, 309 (1998).

\bibitem{Ref6} U. Geiser {\it et al}., Inorg. Chem. {\bf 30}, 2586 (1991).

\bibitem{Ref7} T. Komatsu, N. Matsukawa, T. Inoue, and G. Saito, J. Phys. Soc. Jpn. {\bf 65}, 1340 (1996).

\bibitem{Ref8} U. Welp {\it et al}., 
Phys. Rev. Lett. {\bf 69}, 840 (1992).

\bibitem{Ref9} K. Miyagawa, K. Kawamoto, Y. Nakazawa, and K. Kanoda, 
Phys. Rev. Lett. {\bf 75}, 1174 (1995).

\bibitem{Ref10} J. M. Williams {\it et al}., Inorg. Chem. {\bf 29}, 3272 (1990).

\bibitem{Ref11} N. Elstner, R.R.P. Singh, and A. P. Young, Phys. Rev. Lett. {\bf 71}, 1629 (1993).

\bibitem{Ref12} M. Tamura and R. Kato, J. Phys. Cond. Mat. {\bf 14}, L729 (2002).

\bibitem{Ref13} K. Miyagawa, A. Kawamoto, K. Uchida, K. Kanoda, Physica B {\bf 284-288}, 1589 (2000).

\bibitem{Ref14}	Y. Kitaoka, private communication; Y. Kitaoka, {\it et al}., J. Phys. Soc. Jpn. {\bf 67}, 3703 (1998).

\bibitem{Ref15}	S. Ohira, private communication.

\bibitem{Ref16}	M. Imada, J. Phys. Soc. Jpn. {\bf 56}, 881 (1987).

\bibitem{Ref17}	Z. Hiroi, {\it et al}., J. Phys. Soc. Jpn. {\bf 70}, 3377 (2001).

\bibitem{Ref18} D. A. Huse and V. Elser, Phys. Rev. Lett. {\bf 60}, 2531 (1988); B. Bernu , P. Lecheminant, C. Lhuillier and L. Pierre, Phys. Rev. B {\bf 50}, 10048 (1994).

\bibitem{Ref19}	A.E. Trumper, Phys. Rev. B {\bf 60}, 2987 (1999); L.O. Manuel and H.A. Ceccatto {\it ibid}. {\bf 60}, 9489 (1999); Z. Weihong, R. H. McKenzie, R. P. Singh, {\it ibid}. {\bf 59}, 14367 (1999); J. Merino {\it et al}., J. Phys.: Condens. Matter {\bf 11}, 2965 (1999); C.H. Chung {\it et al}., {\it ibid} {\bf 13}, 5159 (2001).

\bibitem{Ref20}	 R. Coldea et al. Phys. Rev. Lett. {\bf 86}, 1335 (2001).

\bibitem{Ref21}	H. Morita, S. Watanabe, M. Imada, J. Phys. Soc. Jpn. {\bf 71}, 2109 (2002).

\bibitem{Ref22}	S. Dual and D. J. Scalapino, Phys. Rev. B {\bf 62}, 8658 (2000).

\bibitem{Ref23}	K. Miyagawa, A. Kawamoto, K. Kanoda, Phys. Rev. Lett. {\bf 89}, 17003 (2002).

\end{thebibliography}
\end{document}